\documentclass{jpsj3}

\title{Hamiltonian Determination with Restricted Access in Transverse Field 
Ising Chain}
\author{
Mohammad Ali Fasihi$^1$,
Shu Tanaka$^{2}$\footnote{E-mail: shu-t@alice.math.kindai.ac.jp},
Mikio Nakahara$^{1,2}$,
and
Yasushi Kondo$^1$
}

\inst{
$^1$Department of Physics, Kinki University, 3-4-1 Kowakae, Higashi-Osaka, Osaka 577-8502, Japan.
$^2$Research Center for Quantum Computing, Interdisciplinary Graduate School of Science and Engineering, Kinki University, 3-4-1 Kowakae, Higashi-Osaka, Osaka 577-8502, Japan.
}

\abst{
We propose a method to evaluate parameters in the Hamiltonian of
the Ising chain under site-dependent transverse fields, with a
proviso that we can control and measure one of the edge spins only. 
We evaluate the eigenvalues
of the Hamiltonian and the time-evoultion operator exactly for a 3-spin chain,
from
which we obtain the expectation values of $\sigma_x$
of the first spin.
The parameters are found from the peak positions of the Fourier transform
of the expectation value.
There are four assumptions in our method, which are mild enough
to be satisfied in many physical systems.
}

\kword{Hamiltonian determination, Quantum dynamics, 
Transverse field Ising chain}

\begin{document}
\maketitle

\section{Introduction}

Quantum information processing has been studied for a long time since 
Feynman's pioneering work\cite{Shor-1994,Nielsen-2000,Nakahara-2008}.
Many researchers are working toward physical realizations of a practical
quantum computer.
We have to overcome several obstacles, however, to physically
realize a working quantum information processor. One of the main issues 
is to synthesize a high-quality qubit with a long coherence time.
If we prepare a set of high-quality qubits, we have to identify 
the interaction strengths among qubits with a high degree of accuracy.
Without knowing the inter-qubit interactions, it is impossible to control 
the qubit state at our disposal. In some cases, the majority of the
 qubits must be isolated from the environment 
to suppress decoherence, in which case we have an access to a 
single qubit only. Then we need to find the parameters in the
total system, such as the coupling constants, through the single qubit.

There are existing works on evaluation of spin-spin interactions with such a 
restricted access\cite{Mohseni-2006,Kohout-2008,Bendersky-2008,Burgarth-2009a,Burgarth-2009b,Burgarth-2010,Shikano-2010}. 
Burgarth, Maruyama and Nori\cite{Burgarth-2009a,Burgarth-2009b,Burgarth-2010},
studied the Hamiltonian determination in the Heisenberg and the XXZ models
with $N$ spins.
They took advantage of the fact that the excited states in these systems are 
classified according 
to the eigenvalues of a conserved quantity and each excited state belongs to
an $N$-dimensional subspace of the total Hilbert space. As a result, it is
sufficient to work with this
$N$-dimensional subspace in evaluating the parameters in the Hamiltonian.

The purpose of this paper is to extend their analysis to more challenging
cases in which the whole $2^N$ basis vectors are required to construct
an eigenvector. Our particular example studied here is the Ising chain
in site-dependent transverse fields. We restrict ourselves mostly within
a three-spin chain, the shortest nontrivial chain, to make our 
analysis concrete. It also turns out that
the result is extremely lengthy for a longer spin chain so that it is
impossible to write down the results in compact forms. 

We introduce our model Hamiltonian and quantum dynamics in the next section.
Section 3 is devoted to Hamiltonian determination. 
Discussion and conclusion are given in Section 4.
Details of the calculation are summarized in Appendix.

\section{Hamiltonian and Time-Evolution}

We investigate, in this paper, how to determine the spin-spin interaction
strengths and the external magnetic field strengths in a $3$-spin Ising 
chain with site-dependent transverse fields. The Hamiltonian is given by
\begin{eqnarray}\label{eq:hamiltonian}
 {\cal H}(h_i, J_i) = \sum_{i=1}^{2} J_i \sigma_i^z \sigma_{i+1}^z
  - \sum_{i=1}^3 h_i \sigma_i^x,
\end{eqnarray}
where $\sigma_i^x$ and $\sigma_i^z$ denote the $x$-component and the
$z$-component of the Pauli spin matrices at the $i$-th site, respectively.
It is assumed, without loss of generality, 
that the magnetic field is applied along the $x$-axes of spins 2 and 3
by employing local gauge transformations.
This is due to the fact that the dynamics of $\sigma_1^x$
is independent of the direction of the magnetic fields $h_2$ and $h_3$
in the $xy$-plane at the sites 2 and 3, respectively. 
This can be explicitly seen by the identity
\begin{equation}
V^{\dagger}\sigma_1^x V= \sigma_1^x,
\end{equation}
where $V = e^{-i (\alpha_2 \sigma_2^z + \alpha_3 \sigma_3^z)}$
is a transformation which rotates the spins 2 and 3 in the $xy$-plane
by angles $\alpha_2$ and $\alpha_3$, respectively. 
We take $h_2$ and $h_3$ to be positive by making use of this freedom from
now on. 

Our purpose is to evaluate the parameters in the Hamiltonian by controlling
and measuring the first spin only. We assume the following four mild 
conditions to this end:
 \begin{itemize}
  \item It is possible to prepare the fully polarized state
$| \uparrow \uparrow  \uparrow \rangle$ as an initial state.
  \item We can control the first spin only. We may prepare the states such 
as $(\alpha |\uparrow \rangle + \beta |\downarrow \rangle)
| \uparrow  \uparrow \rangle $ by making use of this assumption.
  \item Once the initial state is set up, we can control the magnetic 
field $h_1$ at the first site only. Other fields
$h_2$ and $h_3$ are fixed but unknown to us.
  \item We can measure the spin dynamics of $\sigma_1^x$ only.
 \end{itemize}
%

Our proposed scheme is summarized as follows:
\begin{description}
 \item[Step 1] We prepare an initial state.
 \item[Step 2] The dynamics of the $x$-component of the first 
spin $\langle \sigma_1^x (t) \rangle$ is measured.
 \item[Step 3] The Fourier transform $\hat{\sigma}_1^x(\omega)$
of $\langle \sigma_1^x (t) \rangle$ is evaluated.
 \item[Step 4] We determine $h_2, h_3$, $|J_{1}|$ and $|J_{2}|$ from the peak 
positions of $\hat{\sigma}_1^x (\omega)$ as functions of $h_1$.
 \item[Step 5] There are four combinations of the signs of $J_1$ and $J_2$.
We evaluate the short time behavior of $\sigma_1^x(t)$ numerically for each
of the four cases and compare the results with that obtained
experimentally to determine the signs.
(The dynamics is independent of the signs of $h_2$ and $h_3$ 
as remarked previously.)
\end{description}
%

Let $\{\epsilon_i\}_{1 \leq i \leq 8}$ be the eigenenergies of the Hamiltonian, whose explicit
forms in terms of the parameters $\{h_i, J_i\}$ are given by 
Eq.~(\ref{eq:eigenenergies}) in Appendix.
The eigenenergies of this Hamiltonian satisfy the following symmetry relations:
\begin{eqnarray}
 \begin{cases}\label{eq:relation1}
  \epsilon_5 = - \epsilon_4 \equiv \epsilon_{\bar{4}}\\
  \epsilon_6 = - \epsilon_3 \equiv \epsilon_{\bar{3}}\\
  \epsilon_7 = - \epsilon_2 \equiv \epsilon_{\bar{2}}\\
  \epsilon_8 = - \epsilon_1 \equiv \epsilon_{\bar{1}}
 \end{cases}
\end{eqnarray}
\begin{eqnarray}
 \begin{cases}
  \epsilon_1 + \epsilon_4 = \epsilon_2 + \epsilon_3 = -(\epsilon_{\bar{1}}+
\epsilon_{\bar{4}}
)= -(\epsilon_{\bar{2}}+\epsilon_{\bar{3}}) \\
  \epsilon_1 - \epsilon_3 = \epsilon_2 - \epsilon_4 = -(\epsilon_{\bar{1}}-
\epsilon_{\bar{3}})
 = -(\epsilon_{\bar{2}}-\epsilon_{\bar{4}}) \\
  \epsilon_1 - \epsilon_2 = \epsilon_3 - \epsilon_4=
-(\epsilon_{\bar{1}} -\epsilon_{\bar{2}})
=-(\epsilon_{\bar{3}}- \epsilon_{\bar{4}})
 \end{cases}
\end{eqnarray}

Let us choose the fully polarized state 
$|\uparrow \uparrow \uparrow \rangle$ as an initial state.
The state of the chain at time $t>0$ is $|\psi(t) \rangle = U(t)
|\uparrow \uparrow \uparrow \rangle$, where $U(t)$ is
the time evolution operator 
\begin{eqnarray}
U(t) &=& \exp(-i {\mathcal H}t)
= \sum_{j=1}^{8} e^{- i \epsilon_j t} {\mathcal P}_j.
\end{eqnarray}
Here
\begin{equation}
{\mathcal P}_j = \prod_{\substack{k=1 \\ k \neq j}}^8 
\frac{{\mathcal H}-\epsilon_k I}{\epsilon_j - \epsilon_k} 
\end{equation}
is the projection operator to the eigenspace with the eigenenergy
$\epsilon_j$.
The real-time dynamics $\langle \sigma_1^x (t) \rangle$ of the first spin 
is calculated with respect to $|\psi(t) \rangle$ as
\begin{eqnarray}\label{eq:sigma1xt}
 \left\langle \sigma_1^x (t) \right\rangle
= \langle \psi(t)|\sigma_1^x   |\psi(t) \rangle 
  = C+\sum_{\substack{m=1\\ m > n}}^4 
  \left[
   A_{mn} \cos \left( \epsilon_m + \epsilon_n \right) t + 
   B_{mn} \cos \left( \epsilon_m - \epsilon_n \right) t
  \right],
\end{eqnarray}
where the coefficients $C$,
$A_{mn}$ and $B_{mn}$ 
are functions of $\{ J_i \}$ and $\{ h_i \}$. For example, 
$C$ is explicitly written as
\begin{equation}
C = \langle \uparrow \uparrow \uparrow |
\sum_{i=1}^8 {\mathcal P}_i \sigma_1^x {\mathcal P}_i| \uparrow\uparrow\uparrow
\rangle,
\end{equation}
while $A_{21}$ and $B_{41}$ are
\begin{equation}
A_{21} = \langle \uparrow \uparrow \uparrow |\left(
{\mathcal P}_1 \sigma_1^x {\mathcal P}_7 
+{\mathcal P}_2 \sigma_1^x {\mathcal P}_8
+{\mathcal P}_7 \sigma_1^x {\mathcal P}_1
+{\mathcal P}_8 \sigma_1^x {\mathcal P}_2\right)| \uparrow\uparrow\uparrow
\rangle,
\end{equation}
and
\begin{equation}
B_{41} = \langle \uparrow \uparrow \uparrow |\left(
{\mathcal P}_1 \sigma_1^x {\mathcal P}_4 
+{\mathcal P}_4 \sigma_1^x {\mathcal P}_1
+{\mathcal P}_8 \sigma_1^x {\mathcal P}_5
+{\mathcal P}_5 \sigma_1^x {\mathcal P}_8\right)| \uparrow\uparrow\uparrow
\rangle,
\end{equation}
respectively.

It is found from Eq.~(\ref{eq:sigma1xt}) that 
the peak positions of the Fourier transform $\hat{\sigma}_1^x (\omega)$ 
are {\it potentially}
at $\epsilon_m \pm \epsilon_n$ for all combinations of $m$ and $n$ 
($1\leq m,n \leq 4, m \neq n$). It should be noted, however, that
some combinations of $m$ and $n$ have vanishing amplitudes 
$A_{mn}^x, B_{mn}^x$, for which case the corresponding peaks do not exist.
In the present case, it can be shown exactly that 
\begin{equation}
A_{41}^x= A_{32}^x = B^x_{21}=B^x_{31}=B^x_{42}=B^x_{43}=0.
\end{equation}

These amplitudes satisfy the sum rule
\begin{eqnarray}\label{eq:sumrule1}
C+ \sum_{\substack{m=1\\ m > n}}^4 (A_{mn} + B_{mn})
&=& \langle \uparrow \uparrow \uparrow|\left(\sum_{j=1}^8 {\mathcal P}_j
\right) \sigma_1^x \left(\sum_{k=1}^8 {\mathcal P}_k \right) |
\uparrow \uparrow \uparrow \rangle\nonumber \\
&=& \langle \uparrow \uparrow \uparrow| \sigma_1^x |
\uparrow \uparrow \uparrow \rangle = 0,
\end{eqnarray}
where use has been made of the completeness relation
$$
\sum_{j=1}^8 {\mathcal P}_j =I.
$$

\section{Hamiltonian Determination}

\begin{figure}[h]
 \begin{center}
  \includegraphics[scale=1]{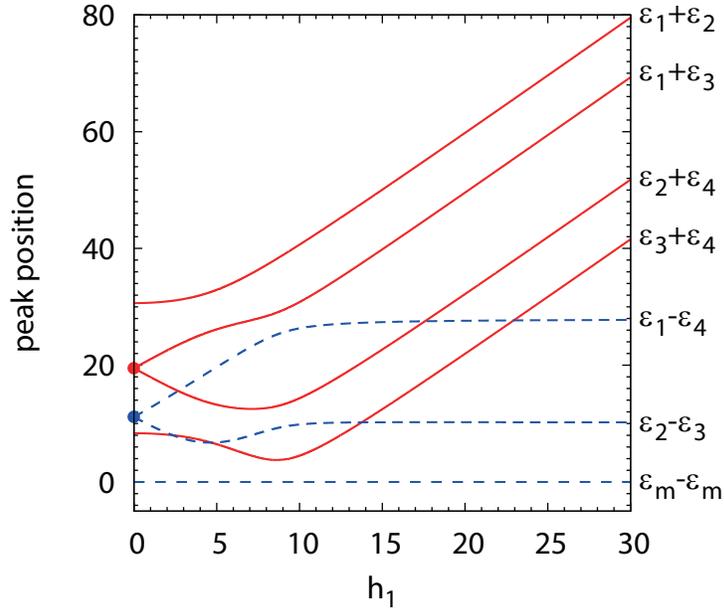}
  \caption{(Color online) Red solid and blue broken curves denote $\epsilon_m + \epsilon_n$
and $\epsilon_m - \epsilon_n$, respectively,
as functions of the transverse field $h_1$ with $h_2=6$, $h_3=7$, $J_1=4$ 
and $J_2=5$.
We have plotted the first quadrant part of the graph
for simplicity. The peaks in the other quadrants are obtained by mirror 
reflections of this diagram with respect to the horizontal and/or the
vertical axes. The red (blue) dot at $h_1=0$ shows
the nonvainshing amplitude $A_{\rm II,I}^{0}$ ($B_{\rm II,I}^{0}$). 
}
  \label{graph:engsumdiff}
 \end{center}
\end{figure}

Suppose we conducted an experiment and measured 
$\left\langle \sigma_1^x (t) \right\rangle$,
or equivalently $\hat{\sigma}_1^x (\omega)$.
Now our task is to identify which peak corresponds to
which $\epsilon_m \pm \epsilon_n$ in order to evaluate the parameters $\{J_i\}$
and $\{h_i\}$.
Figure \ref{graph:engsumdiff} shows $\epsilon_m \pm \epsilon_n$ with
non-vanishing $A_{mn}$ and $B_{mn}$
as functions of the transverse 
field $h_1$, which is assumed to be controllable.
Other parameters are arbitrarily
chosen as $h_2=6$, $h_3=7$, $J_1=4$ and $J_2=5$.
The red solid curves and blue broken
 curves in Fig.~\ref{graph:engsumdiff} depict
$\epsilon_m + \epsilon_n$ and $\epsilon_m - \epsilon_n$, respectively.
It should be clear from this diagram that
the curves cross intricately for intermediate $h_1$, while the behaviors
of the curves for small $h_1$ and large $h_1$ are regular. Then identification
of the curves and $\epsilon_m \pm \epsilon_n$ in these regions is easy.
It should be noted here that the case $h_1=0$ is singular:
$\sigma_1^x(t)$ remains zero for any $t \geq 0$ for $h_1=0$
and $|\psi(0) \rangle = |\uparrow \uparrow \uparrow \rangle$.
This case will be treated separately below.

\begin{figure}[h]
 \begin{center}
  \includegraphics[width=13cm]{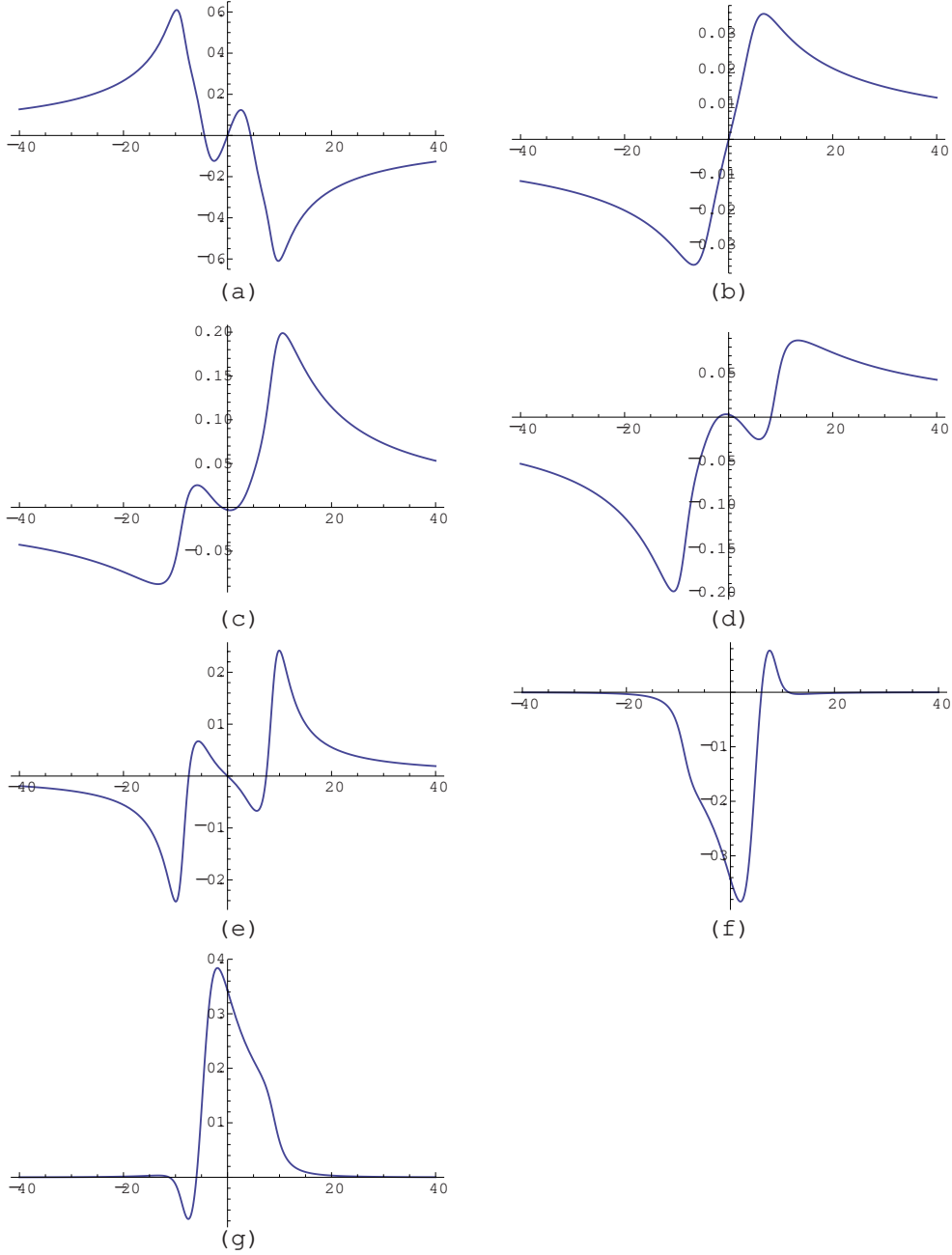}
  \caption{(Color online) Peak amplitudes (a) $C$, (b) $A_{21}$, (c) $A_{31}$, (d) $A_{42}$, 
(e) $A_{43}$, (f) $B_{41}$ and (g) $B_{32}$ as functions of $h_1$. Other
parameters are $h_2=6$, $h_3=7$, $J_1=4$ and $J_2=5$.
All other amplitudes vanish identically.
}
  \label{graph:aandb}
 \end{center}
\end{figure}

In principle,
we may also use the heights (amplitudes) $A_{mn}$ and $B_{mn}$ of theses peaks
to determine the parameters. In practice, however, it is rather
difficult to use the data for this purpose due to the following reason.
First, the analytical expressions for $A_{mn}$ and $B_{mn}$ are extremely
lengthy and we need to employ extensive numerical optimization to determine
the parameters which fit with the data. Second, the peak has finite width 
in actual experiment due to various reasons, such as relaxation.
We must integrate the peak
profile with respect to $\omega$ to find the amplitude, which always 
suffers from error. In addition, reading off the frequency can be done with
much better precision than measuring the amplitude. Figure~\ref{graph:aandb}
shows
non-vanishing amplitudes as functions of $h_1$ for $h_2=6$, $h_3=7$, 
$J_1=4$ and $J_2=5$.
Peak amplitudes $C$, $A_{21}$ and $A_{43}$ are odd functions of $\omega$ while
others are apparently not. Note, however, that the positive-$h_1$ 
(negative-$h_1$) part
of $A_{31}$ and the negative-$h_1$ (positive-$h_1$) part of $A_{42}$ combined
together define an odd function. The loss of antisymmetry in these amplitudes
is due to our choice of the energy eigenvalues (\ref{eq:eigenenergies}).
If we demand that the eigenvalues be defined uniquely by 
Eq.~(\ref{eq:eigenenergies}) for all parameters, we must give up the
antisymmetry of the amplitudes $A_{mn}$ and $B_{mn}$. This also happens
to the pair $B_{41}$ and $B_{32}$.
Figure~\ref{graph:aandb} demonstrates that the sum rule (\ref{eq:sumrule1})
holds for any $h_1$.

Let us first consider the limit of 
a strong magnetic field $h_1$, that is, the parameters satisfy
$|h_1| \gg h_2, h_3, |J_1|, |J_2|$.
The non-vanishing 
Fourier peak positions $\epsilon_m + \epsilon_n$ $(1 \leq m,n \leq 4)$
diverge linearly as $h_1 \to \infty$.
In contrast, the peak positions $\epsilon_m - \epsilon_n$ converge
to constant values.
We obtain the asymptotic behavior of $\epsilon_m \pm \epsilon_n$
for $|h_1| \gg h_2, h_3, |J_1|, |J_2|$ from  
the explicit forms of $\epsilon_m$ given in Eq.~(\ref{eq:eigenenergies}) as 
\begin{eqnarray}
 &&
 \epsilon_1 + \epsilon_2
  \to 2h_1 + \left[
	    \sqrt{\left( h_2 + h_3 \right)^2 + J_2^2} + \sqrt{\left( h_2 - h_3 \right)^2 + J_2^2}
	   \right]
  + {\cal O}\left( \frac{1}{h_1}\right),\label{eq:1p2}\\
 &&
\epsilon_1 + \epsilon_3 
  \to 2h_1 + \left[
	    \sqrt{\left( h_2 + h_3 \right)^2 + J_2^2} - \sqrt{\left( h_2 - h_3 \right)^2 + J_2^2}
	   \right]
  + {\cal O}\left( \frac{1}{h_1}\right),\\
 &&
\epsilon_2 + \epsilon_4 
  \to 2h_1 - \left[
	    \sqrt{\left( h_2 + h_3 \right)^2 + J_2^2} - \sqrt{\left( h_2 - h_3 \right)^2 + J_2^2}
	   \right]
  + {\cal O}\left( \frac{1}{h_1}\right),\\
 &&
\epsilon_3 + \epsilon_4 
  \to 2h_1 - \left[
	    \sqrt{\left( h_2 + h_3 \right)^2 + J_2^2} + \sqrt{\left( h_2 - h_3 \right)^2 + J_2^2}
	   \right]
  + {\cal O}\left( \frac{1}{h_1}\right)\label{eq:3p4}
\end{eqnarray}
and
\begin{eqnarray}
 &&
\epsilon_1 - \epsilon_4 
  \to 2\sqrt{\left( h_2 + h_3 \right)^2 + J_2^2}
  + {\cal O}\left( \frac{1}{h_1^2}\right) ,\\
 &&
\epsilon_2 - \epsilon_3 
   \to 2\sqrt{\left( h_2 - h_3 \right)^2 + J_2^2}
   + {\cal O}\left( \frac{1}{h_1^2}\right).
\end{eqnarray}
From these results, we find the ordering of the peak positions for a 
large $h_1$ as
\begin{equation}
\epsilon_1 + \epsilon_2 > \epsilon_1 + \epsilon_3 > \epsilon_2 + \epsilon_4 >
\epsilon_3 + \epsilon_4   
\end{equation}
and
\begin{equation}
\epsilon_1 - \epsilon_4 > \epsilon_2 - \epsilon_3.
\end{equation}
Once the peak positions are identified with particular
$\epsilon_m \pm \epsilon_n$, we may easily find 
$\sqrt{\left( h_2 + h_3 \right)^2 + J_2^2}$
and $\sqrt{\left( h_2 - h_3 \right)^2 + J_2^2}$ numerically.
Clearly these two numbers are not sufficient to determine the
four parameters in the Hamiltonian and we need to examine the
peaks at other values of $h_1$, such as $h_1=0$, to acquire
sufficient data to completely determine the parameters.

We note {\it en passant} that
the asymptotic behavior of the peaks does depend on $J_2$ but not on $J_1$.
This counter-intuitive character is understood if we study the perturbation 
theory at $h_1 \to \infty$. Let us separate the Hamiltonian as
\begin{eqnarray}
 &&{\cal H} = {\cal H}_0 + {\cal H}_{\rm 1},\\
 &&{\cal H}_0 = - h_1 \sigma_1^x, \\
 &&{\cal H}_{\rm 1} = \sum_{i=1}^{2} J_i \sigma_i^z \sigma_{i+1}^z - \sum_{i=2}^3 h_i \sigma_i^x.
\end{eqnarray}
The eigenvectors of the unperturbed Hamiltonian ${\cal H}_0$ are 
$\frac{1}{\sqrt{2}} (|\!\uparrow \rangle \pm |\!\downarrow \rangle)
|s_2 s_3 \rangle$, where $s_2, s_3 \in \{\uparrow,
\downarrow\}$.
The first order perturbation of the eigenvalues is independent of $J_1$, which
is why the asymptotic eigenenergies are independent of $J_1$.
This behavior is also understood from the spin dynamics.
When the magnetic field $h_1$ is much larger than the 
other parameters, the characteristic time $\tau$ of the motion of the first 
spin is very short compared to other characteristic times.
Then the effect of the spin-spin interaction $J_1$ is averaged to vanish over 
the time interval $t \gg \tau$ and the $J_1$ interaction
is ignored in 
this limit. Clearly this remains true for any number of spins in the chain.

Next we consider the case when $h_1 = 0$ to acquire sufficient data to fix all
the parameters. The eigenenergies in this case are
$\pm \epsilon_{\rm I}, \pm \epsilon_{\rm II}$, each of which is doubly
degenerate. Here we take $\epsilon_{\rm I} > \epsilon_{\rm II}>0$.
The explicit forms of $\epsilon_{\rm I,II}$ are
given by Eq.~(\ref{eq:e0II}) in Appendix.
It is easy to show that
$$
\langle \uparrow \uparrow \uparrow |U(t)^{\dagger} \sigma_1^x
U(t)|\uparrow \uparrow \uparrow \rangle = 0
$$
for $h_1=0$. This is an artifact of the initial state chosen above.
Note that $\sigma_1^x$ does not commute with the Hamiltonian and a
proper initial condition introduces a nontrivial expectation value
of the observable.
Let us take an initial state $|+ \uparrow \uparrow \rangle$, where
$$
|+ \rangle = \frac{1}{\sqrt{2}}(|\!\uparrow \rangle+ |\!\downarrow \rangle).
$$
Then the real-time dynamics $\langle \sigma_1^x (t) \rangle$ is calculated as
\begin{eqnarray}\label{eq:h10}
 \left\langle \sigma_1^x (t) \right\rangle
  = C + 
A_{\rm II,I}^0 \cos \left( \epsilon_{\rm I} + \epsilon_{\rm II} \right) t + 
  B_{\rm II,I}^{0} \cos \left( \epsilon_{\rm I} - \epsilon_{\rm II} \right) t,
\end{eqnarray}
where the coefficients $C$,
$A_{\rm II,I}^0$ and $B_{\rm II,I}^0$ are functions of 
$\{ J_i \}$ and $\{ h_i \}$, whose explicit forms 
are given in Eq.~(\ref{eq:h10}).

The peak positions with nonvanishing amplitudes
at $h_1=0$ are
\begin{eqnarray}
 \label{eq:1p3_0}
 && \omega = \epsilon_{\rm I} \pm \epsilon_{\rm II}
 = \sqrt{S_0 + 2 \sqrt{C_0}} \pm \sqrt{S_0 - 2 \sqrt{C_0}},
\end{eqnarray}
and $\omega = 0$,
%
%
where
\begin{eqnarray}
 S_0 = h_2^2 + h_3^2 + J_1^2 + J_2^2,\ C_0 = h_2^2 h_3^2 + h_3^2 J_1^2 + J_1^2 J_2^2.
\end{eqnarray}
Since peak positions $\epsilon_{\rm I} \pm \epsilon_{\rm II}$
produce only two
numbers $\sqrt{S_0 \pm 2 \sqrt{C_0}}$, which
are essentially the same as $S_0$ and $C_0$, 
we should combine the data acquired for the case of a large $h_1$ to 
determine all the parameters $h_2$, $h_3$, $|J_1|$ and $|J_2|$. %
Once all the non-vanishing peak positions
are identified, the measurement of the peaks
of $\hat{\sigma}_1^x(\omega)$ for additional finite $h_1$ helps us
determine the parameters with higher precision.
Figure~\ref{graph:aandb} shows that all the peak
amplitudes decay as
$h_1 \to \infty$. It may happen that the peaks are not observable
at large enough $h_1$ since their amplitudes vanish in this
region. If this is the case, we need to identify peaks at
$h_1=0$ first and then sweep $h_1$ to acquire sufficient data to
determine the parameters. 

For the parameter choice, $h_1=0, h_2=6, h_3=7, J_1=4$ and $J_2=5$, we
obtain the amplitudes numerically as
\begin{equation}
C= 0.5984,\ A_{\rm II,I}^0= 0.05525 ,\ B_{\rm II,I}^0= 0.3464.
\end{equation}
They do not agree with the $h_1 \to 0$ limit in Fig.~\ref{graph:aandb} since
their initial states are different.
These amplitudes satisfy a sum rule different from Eq.~(\ref{eq:sumrule1}),
\begin{equation}
C+A_{\rm II,I}^0 + B_{\rm II,I}^0 = \langle + \uparrow \uparrow |
\sigma_1^x |+ \uparrow \uparrow 
\rangle = 1.
\end{equation}

\begin{figure}[h]
 \begin{center}
  \includegraphics[scale=0.7]{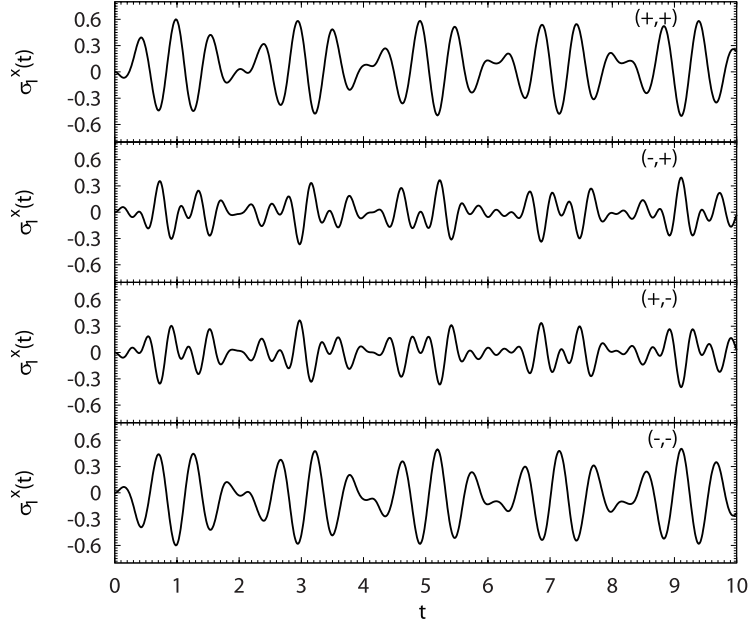}
  \caption{
  Dynamics of $\sigma_1^x(t)$ for $({\rm sign}(J_1), {\rm sign}(J_2))=(+,+)$, 
$(-,+)$, $(+,-)$, and $(-,-)$, where the Hamiltonian parameters are 
$h_1=1$, $h_2=6$, $h_3=7$, $|J_1|=4$ and $|J_2|=5$.
  }
  \label{graph:sign}
 \end{center}
\end{figure}

Now that  $h_2$, $h_3$, $|J_1|$ and $|J_2|$ are determined,
our next task is to find the signs of $J_1$ and $J_2$.
Figure \ref{graph:sign} shows the dynamics of $\sigma_1^x$ 
for $|J_1|=4$, $|J_2|=5$, $h_1=1$, 
$h_2=6$, and $h_3=7$ for the four combinations of the signs of 
$J_1$ and $J_2$. 
The signs are $({\rm sign}(J_1), {\rm sign}(J_2))=(+,+)$, 
$(-,+)$, $(+,-)$, and $(-,-)$ from top to bottom.
We immediately note that $\sigma_1^x(t)$ is positive (negative)
for small time limit
if ${\rm sign}(J_1)$ is negative (positive). 
This is because the molecular field at the first site is approximately
$(-h_1, 0, J_1)$
for small $t$. Thus sign($J_1$) is fixed by inspecting
the short time dynamics
of $\langle \sigma_1^x (t)\rangle$. Unfortunately, sign($J_2$) cannot
be determined in this way. We also note
that the dynamics $\langle \sigma_1^x (t) \rangle$ 
changes the sign under the transformation $(J_1, J_2) \to (-J_1, -J_2)$
as seen from Fig.~\ref{graph:sign}. This is due to the following
identities,
\begin{equation} 
X_{2,\pi}(-\sum_{i=1}^3 h_i \sigma_i^x + \sum_{i=1}^2 J_i \sigma_i^z
\sigma_{i+1}^z) X_{2,\pi}^{\dagger}  = 
-\sum_{i=1}^3 h_i \sigma_i^x - \sum_{i=1}^2 J_i \sigma_i^z
\sigma_{i+1}^z
\end{equation}
and
\begin{eqnarray}
\lefteqn{ 
\langle \uparrow \uparrow \uparrow |e^{i {\mathcal H}(h_i, -J_i)t}
\sigma_1^x e^{-i {\mathcal H}(h_i, -J_i)t}| \uparrow \uparrow  \uparrow 
\rangle }\nonumber\\
&=&  
\langle \uparrow \uparrow \uparrow |X_{2,\pi}^{\dagger}
e^{i {\mathcal H}(h_i, J_i)t}X_{2,\pi}
\sigma_1^x X_{2,\pi}^{\dagger}e^{-i {\mathcal H}(h_i, J_i)t}X_{2,\pi}
| \uparrow \uparrow  \uparrow \rangle \nonumber\\
&=& - \langle \uparrow \uparrow \uparrow |
e^{i {\mathcal H}(h_i, J_i)t} \sigma_1^x e^{-i {\mathcal H}(h_i, J_i)t}
| \uparrow \uparrow  \uparrow \rangle,
\end{eqnarray}
where $X_{2,\pi} = \exp(-i \sigma_2^x \pi/2)$ is a unitary matrix corresponding
to a rotation of the spin 2 around the $x$-axis by an angle $\pi$.

\section{Summary and Discussion}

In this paper, we propose a method to evaluate the spin-spin interaction 
strengths and the site-dependent magnetic field strengths only through 
one of the edge spins. We consider a three-spin Ising model with
site-dependent transverse fields $h_i$.
Since the Ising model with transverse fields has no good quantum number 
such as the total $S^z$, which the Heisenberg and the XXZ models have,
it is considerably 
more difficult to evaluate the parameters in the Hamiltonian 
in this case.

First a fully polarized state $|\uparrow \uparrow \uparrow \rangle$ 
(when $h_1 \neq 0$) or $\frac{1}{\sqrt{2}}(|\uparrow \rangle
+ |\downarrow \rangle)|\uparrow \uparrow \rangle$ (when $h_1=0$)
is prepared 
and the time-dependence of the $x$-component of the first spin
is measured. The peak positions of the Fourier
transform $\hat{\sigma}_1^x(\omega)$ are at $\{\epsilon_m \pm \epsilon_n \}$,
where $\epsilon_n$ is the first four eigenvalues of the Hamiltonian.
The identification of each peak position with a particular
$\epsilon_m \pm \epsilon_n$
is easy for $h_1=0$ and $h_1 \to \infty$, while it is difficult for a finite
$h_1$ in general. Of course, it is possible to identify the peaks for a
finite $h_1$ if $h_1$ is swept from $0$ to a large value continuously.

We determine $\{h_i\}$ and $\{|J_i|\}$ completely from the peak positions
of $\sigma_1^x(\omega)$.
It was shown that some possible peaks are
not observable since they have vanishing amplitudes.
Some combinations of the parameters are read off from the peak positions
at large $h_1$ and $h_1=0$.
The combined
data is sufficient to fix all the parameters completely. Additional 
data obtained
for finite $h_1$ improves accuracy of the estimated parameters.
The signs of the spin-spin interactions are found by comparing the 
experimentally obtained spin dynamics with numerical simulation.

In our proposed method, there are only four mild conditions which are
satisfied in many experimental systems such as a liquid state
NMR quantum computer, which is abbreviated as ``an NMR quantum computer''
hereafter. 
In fact, the Ising model with site-dependent transverse field is a typical 
system in an NMR quantum computer. 
It is possible to control the magnetic field strength $h_i$
of each spin independently by making coordinate transformation to a 
rotating frame of each spin rotating with the respective Lamour frequency,
It should be noted, however, that there is a subtle problem
in an NMR quantum computer associated with the initial state.
The initial pure state is prepared as a pseudo-pure state,
which is produced by the time-average or the space-average
method, for example.\cite{pps-1,pps-2,pps-3,pps-4,pps-5}
To produce a pseudo-pure state, however, we need
to know the Hamiltonian in advance, which is ``the chicken or the 
egg'' causality dilemma. The best we can do is to prepare a
pseudo-pure state with a molecule with a known Hamiltonian and
then pretend as if we do not know the Hamiltonian and demonstrate
our method.
We believe our scheme can be experimentally demonstrated also
with other physical systems with a pure state.

The authors are grateful to Daniel Burgarth and Koji Maruyama for
valuable comments and discussions. This work is partially supported by 
`Open Research Center' Project for Private Universities: matching fund 
subsidy from MEXT, Japan.
The authors thank the Yukawa Institute for Theoretical Physics at Kyoto 
University. Discussions during the YITP workshop YITP-W-10-14 on ``Duality 
and Scale in Quantum-Theoretical Sciences'' were useful to complete this work. 
S.T. is partly supported by Grant-in-Aid for Young Scientists Start-up (21840021) from the JSPS and by MEXT Grant-in-Aid for Scientific Research (B) (22340111).
The computation in the present work was performed on computers at the 
Supercomputer Center, Institute for Solid State Physics, University of Tokyo.

\appendix
\section{Derivation of Physical Quantities}

\subsection{Eigenenergies}

After straightforward but tedious calculation,
eigenenergies of the Hamiltonian (\ref{eq:hamiltonian}) 
are obtained explicitly as:
\begin{eqnarray}
 \label{eq:eigenenergies}
  \begin{cases}
   \displaystyle \epsilon_1 = \frac{\alpha}{2} + \frac{\left( \beta + \gamma \right)^{1/2}}{2} \\
  \displaystyle  \epsilon_2 = \frac{\alpha}{2} + \frac{\left( \beta - \gamma \right)^{1/2}}{2} \\
  \displaystyle  \epsilon_3 = \frac{\alpha}{2} - \frac{\left( \beta - \gamma \right)^{1/2}}{2} \\
  \displaystyle  \epsilon_4 = \frac{\alpha}{2} - \frac{\left( \beta + \gamma \right)^{1/2}}{2} \\
  \end{cases}
  \hspace{10mm}
  \begin{cases}
   \epsilon_5 = - \epsilon_4 \equiv \epsilon_{\bar{4}}\\
   \epsilon_6 = - \epsilon_3 \equiv \epsilon_{\bar{3}}\\
   \epsilon_7 = - \epsilon_2 \equiv \epsilon_{\bar{2}}\\
   \epsilon_8 = - \epsilon_1 \equiv \epsilon_{\bar{1}}\\
  \end{cases}
\end{eqnarray}
where
\begin{eqnarray}
& &\alpha = \sqrt{\frac{1}{3} \left( 4S + \frac{B}{A}+\frac{A}{2^{1/3}}\right)},
\ \beta = \frac{1}{3} \left( 8S - \frac{B}{A} - \frac{A}{2^{1/3}} \right),
\ \gamma = \frac{16h_1 h_2 h_3}{\alpha},\nonumber\\
&& S = h_1^2 + h_2^2 + h_3^2 + J_1^2 + J_2^2,
\ A = \left[ 
	X + \sqrt{-256\left( S^2 + 3\sqrt{{\rm det}{\cal H}}\right)^3 + X^2}
       \right]^{1/3},\nonumber \\
&&X = 16\left[ 
	  -S^3 + 9S \sqrt{{\rm det}{\cal H}} + 108 h_1^2 h_2^2 h_3^2
	 \right],
\ B = 2^{7/3} \left[
		S^2 + 3 \sqrt{{\rm det}{\cal H}}
	       \right],\\
&&\sqrt{{\rm det}{\cal H}} = h_1^4 + h_2^4 + \left( h_3^2 - J_1^2 + J_2^2 \right)^2
\nonumber\\
&&\qquad \qquad  - 2h_1^2 \left( h_2^2 + h_3^2 - J_1^2 + J_2^2 \right)
  + 2h_2^2 \left( -h_3^2 + J_1^2 + J_2^2\right).\nonumber
\end{eqnarray}

\begin{figure}[h]
 \begin{center}
  \includegraphics[scale=1]{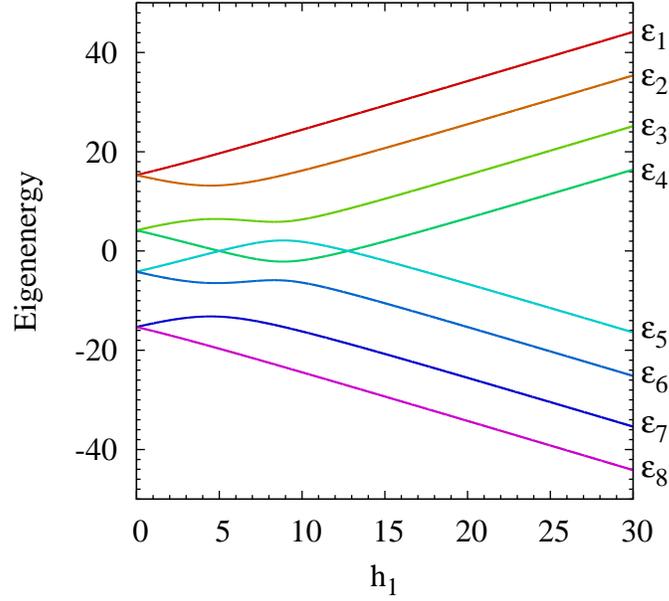}
  \caption{
  (Color online) Eigenenergies as functions of $h_1$ when $J_1=4$, $J_2=5$, 
$h_2=6$, and $h_3=7$.
  }
  \label{graph:eigenenergies}
 \end{center}
\end{figure}

Figure \ref{graph:eigenenergies} shows the eigenenergies as functions of 
the transverse field $h_1$ with $J_1=4$, $J_2=5$, $h_2=6$, 
and $h_3=7$.
All eigeneneries are doubly degenerate when $h_1=0$.

The eigenenergies $\epsilon_4$ and $\epsilon_5$ cross when 
$\epsilon_4=\epsilon_5=0$, which seems to be in contradiction 
with the no-level crossing theorem \cite{lax}. Nonetheless
this is allowed since the eigenvectors corresponding to
these degenerate eigenenergies are symmetric and antisymmetric states
with respect to the spin indices.
The symmetric state and antisymmetric states are characterized by 
introducing the parity operator \cite{Tanaka-2010}:
\begin{eqnarray}
 {\cal X} = \sigma_1^x \otimes \sigma_2^x \otimes \sigma_3^x.
\end{eqnarray}
The Hamiltonian and the parity operator commute:
\begin{eqnarray}
 [{\cal H},{\cal X}] = 0.
\end{eqnarray}
Symmetric and antisymmetric functions are defined as follows:
\begin{eqnarray}
&& |\Psi_{\rm s}\rangle = {\sum_{\left\{ s \right\}}}'
  a_s \left( |s \rangle + {\cal X}|s \rangle \right),\\
&& |\Psi_{\rm as}\rangle = {\sum_{\left\{ s \right\}}}'
  a_s \left( |s \rangle - {\cal X}|s \rangle \right),
\end{eqnarray}
where $|s\rangle=|s_1 s_2 s_3 \rangle$ 
denotes a spin configuration, and $\sum_{\left\{s \right\}}'$ denotes summation over all spin configurations with $s_1 = +1$. Observe that ${\cal X}
|\Psi_{\rm s}\rangle = |\Psi_{\rm s}\rangle$ while ${\cal X}
|\Psi_{\rm as}\rangle = -|\Psi_{\rm as}\rangle$.
The eigenstates with the eigenenergies $\epsilon_4$ and $\epsilon_5$
belong to different symmetry classes and their eigenenergies may
cross as $h_1$ is changed.

\subsection{$h_1=0$ Case}

It was shown in the main text that the measurement of the peak positions
of $\hat{\sigma}_1^x  (\omega)$ at a large $h_1$ was not sufficient
to determine the parameters in the Hamiltonian. 
Let us look at the spin dynamics at $h_1=0$ to obtain information
required to determine the parameters. The initial 
condition $|\uparrow \uparrow\uparrow \rangle$ leads
$\langle \sigma_1^x(t) \rangle= 0$ for any $t$. 
Let us take the initial state $\frac{1}{\sqrt{2}}(|\uparrow
\rangle+|\downarrow \rangle)|\uparrow \uparrow \rangle$l, instead,
to avoid this problem. 

The Hamiltonian is block-diagonalized when $h_1=0$ as
\begin{equation}
{\cal H} = \left(
	\begin{array}{cc}
	 {\cal H}_u & 0 \\
	 0 & {\cal H}_d
	\end{array}
       \right),
\end{equation}
where
\begin{eqnarray}
{\cal H}_{u}
  &=& \left(
     \begin{array}{cccc}
      J_1+J_2 & -h_3 & -h_2 & 0\\
      -h_3 & J_1-J_2 & 0 & -h_2 \\
      -h_2 & 0 & -J_1-J_2 & -h_3 \\
      0 & -h_2 & -h_3 & - J_1 + J_2
     \end{array}
    \right),\\
 {\cal H}_{d}
 & =& \left(
     \begin{array}{cccc}
      -J_1 + J_2 & -h_3& -h_2& 0\\
      -h_3& -J_1-J_2& 0& -h_2\\
      -h_2&0 & J_1-J_2& -h_3\\
      0& -h_2& -h_3& J_1+J_2
     \end{array}
    \right).
\end{eqnarray}
The Hamiltonian has four eigenvalues $\pm \epsilon_{\rm I}, 
\pm \epsilon_{\rm II}$, each of which is doubly degenerate. Here
\begin{equation}
 \epsilon_{\rm I} = \sqrt{S_0 + 2\sqrt{C_0}}, \\ 
\ \epsilon_{\rm II} = \sqrt{S_0 - 2\sqrt{C_0}}  
\label{eq:e0II}
\end{equation}
with
\begin{equation}
S_0 = h_2^2 + h_3^2 + J_1^2 + J_2^2,
\ C_0 = h_2^2 h_3^2 + h_3^2 J_1^2 + J_1^2 J_2^2.
\end{equation}

By using projection operators, time-evolution operators are expressed as
\begin{eqnarray}
 U_{u} &=& {\rm e}^{-i {\cal H}_{u} t}
  = \sum_{k=1}^4 {\cal P}^{(u)}_k {\rm e}^{-i \epsilon_k t}\nonumber\\
  &=& \sum_{\alpha \in \{\rm I, II\}} \left[{\cal P}^{(u)}_\alpha 
{\rm e}^{-i \epsilon_\alpha t} +
  {\cal P}^{(u)}_{\bar{\alpha}} {\rm e}^{-i \epsilon_{\bar{\alpha}} t}\right]
\nonumber\\
 &=& \sum_{\alpha \in \{\rm I, II\}} \left[
{\cal Q}_\alpha^{(u)} \cos (\epsilon_a t) +
{\cal R}_\alpha^{(u)} \sin (\epsilon_\alpha t)\right],\\
 U_{d} &=& {\rm e}^{-i {\cal H}_{d} t}
  = \sum_{k=1}^4  {\cal P}^{(d)}_k {\rm e}^{-i \epsilon_k t} \nonumber\\
  &=& \sum_{\alpha \in \{\rm I, II\}}  \left[
{\cal P}^{(d)}_\alpha {\rm e}^{-i \epsilon_\alpha t} +
  {\cal P}^{(d)}_{\bar{\alpha}} {\rm e}^{-i \epsilon_{\bar{\alpha}} t}
\right]
\nonumber\\
 &=& \sum_{\alpha \in \{\rm I, II\}}  \left[
{\cal Q}_\alpha^{(d)} \cos (\epsilon_a t) +
{\cal R}_\alpha^{(d)} \sin (\epsilon_\alpha t)\right],
\end{eqnarray}
where 
\begin{equation}
{\cal Q}_{\alpha}^{(u,d)}= {\cal P}_{\alpha}^{(u, d)} + 
{\cal P}_{\bar{\alpha}}^{(u, d)},
{\cal R}_{\alpha}^{(u,d)}= {i}\left[{\cal P}_{\alpha}^{(u, d)} +
{\cal P}_{\bar{\alpha}}^{(u, d)}\right].
\end{equation}

Then, real-time dynamics of $x$-component of the first spin is obtained as
\begin{eqnarray}\label{eq:h10}
 \langle \sigma_1^x (t) \rangle
  &=& \frac{h_2^2 h_3^2}{C_0} +A_{\rm II,I}^{0} \cos[(\epsilon_{\rm I}+\epsilon_{\rm II})t] + B_{\rm II,I}^{0}\cos[(\epsilon_{\rm I}-\epsilon_{\rm II})t],
\end{eqnarray}
where
\begin{equation}
\begin{array}{c}
\displaystyle 
A_{\rm II,I}^{0}=\frac{J_1^2 [h_2^2 (h_3^2 - J_2^2) - (h_3^2 + J_2^2) (h_3^2 - J_1^2 + J_2^2)+(J_2^2+h_3^2) \epsilon_{\rm I} \epsilon_{\rm II}]}{2 \epsilon_{\rm I} \epsilon_{\rm II} C_0},\vspace{.3cm}\\
\displaystyle B_{\rm II,I}^{0} = \frac{J_1^2 [h_2^2 (-h_3^2 + J_2^2) + (h_3^2 + J_2^2) (h_3^2 - J_1^2 + J_2^2)+ (J_2^2+h_3^2) \epsilon_{\rm I} \epsilon_{\rm II}]}{2 \epsilon_{\rm I} \epsilon_{\rm II} C_0}.
\end{array}
\end{equation}

\end{document}